\expandafter\ifx\csname natexlab\endcsname\relax\fi

\documentclass{emulateapj}
\slugcomment{{\sc Accepted to ApJ:} May 23rd, 2017} 
\usepackage{amsmath}
\usepackage{mathtools}
\usepackage{hyperref}
\def\a4{\hsize 17.0cm \vsize 25.cm}

\newcommand{\derp}[2] { \frac{\partial #1}{\partial #2} }

\defcitealias{MRN1977}{MRN} 
\defcitealias{MartinezGonzalezetal2016}{MTS16}
\defcitealias{Kroupa2001}{Kroupa}
\shorttitle{}

\shorttitle{The NIR-MIR Excess in Young Massive Clusters}
\shortauthors{Mart\'inez-Gonz\'alez  et al.}

\begin{document}

\title{Can Dust Injected by SN\MakeLowercase{e} Explain the NIR-MIR Excess in Young Massive Stellar Clusters?}

\author{Sergio Mart\'inez-Gonz\'alez \altaffilmark{1},
Richard W\" unsch \altaffilmark{1}, 
Jan Palou\v s \altaffilmark{1} }

\altaffiltext{1}{Astronomical Institute, Academy of Sciences of the Czech Republic, Bo\v cn\' \i \ II 1401-2a, Prague, Czech Republic; martinez@asu.cas.cz}

\begin{abstract}
We present a physically-motivated model involving the different processes affecting supernova dust grains as they
are incorporated into the thermalized medium within young massive star clusters. The model is used to explain the 
near- to mid-infrared (NIR-MIR) excess found in such clusters and usually modeled as a blackbody with temperature 
$\sim (400-1000)$ K. In our approach, dust grains are efficiently produced in the clumpy ejecta of core-collapse 
supernovae, fragmented into small pieces ($\lesssim 0.05$ $\mu$m) as they are incorporated into the hot and dense ISM, 
heated via frequent collisions with electrons and the absorption of energetic photons. Grains with small sizes can more
easily acquire the high temperatures ($\sim 1000$ K) required to produce a NIR-MIR excess with respect to the emission of 
foreground PAHs and starlight. However, the extreme conditions inside young massive clusters make difficult for these 
small grains to have a persistent manifestation at NIR-MIR wavelengths as they are destroyed by efficient thermal 
sputtering. Nevertheless, the chances for a persistent manifestation are increased by taking into account that small 
grains become increasingly transparent to their impinging ions as their size decreases. For an individual SN event, 
we find that the NIR-MIR excess last longer if the time required to incorporate all the grains into the thermalized medium
is also longer, and in some cases, comparable to the characteristic interval between supernova explosions. Our models, 
can successfully explain the near-infrared excesses found in the star clusters observed in M33 \citep{Relanoetal2016} 
assuming a low heating efficiency and mass-loading. In this scenario, the presence of the NIR-MIR excess is 
an indication of efficient dust production in SNe and its subsequent destruction. 
\end{abstract}

\section{Introduction}

It has been extensively shown, both theoretically and observationally, that the ejecta of core-collapse supernovae
(SN) meet the appropriate conditions for the efficient condensation of refractory elements onto massive 
quantities of dust \citep[see for example][]{Cernuschietal1967,Moseleyetal1989,SuntzeffandBouchet1990}. 
From the observational perspective, SN 1987A provides the most compelling evidence so far for the condensation of 
dust in the ejecta of a supernova remnant (SNR), producing $\sim (0.7-0.9)$ M$_{\odot}$ of dust during the 
first $\sim$ 25 years after the explosion of a progenitor star with initial mass $\sim 19$ M$_{\odot}$ 
\citep{Indebetouwetal2014,Matsuuraetal2014}. More recently, \citet{DeLoozeetal2017} and 
\citet{Bevanetal2016} have also derived a large mass of dust condensed in the clumpy ejecta of Cassiopeia A, with 
values ranging from $\sim (0.4-0.6)$ M$_{\odot}$ in the former study and $\sim 1.1$ M$_{\odot}$
in the latter analysis. From the theoretical point of view, one can expect that a single supernova would be able 
to condense $\sim (0.1-3.1)$ M$_{\odot}$ of dust out of progenitors in a mass range 
$\sim (13-80)$ M$_{\odot}$ \citep{TodiniandFerrara2001,Nozawaetal2003}. The presence of inhomogeneities in SN ejecta 
(``clumpiness''), as observed in many SNRs, favors the formation of large quantities of dust \citep{BiscaroCherchneff2016} and 
the coagulation of small grains into larger aggregates \citep{SarangiCherchneff2015}. That is not the end of the story, as 
these dust grains might be heavily eroded by the action of thermal and kinetic sputtering and, in the presence
of turbulence, disrupted in grain-grain collisions as they are reached by the supernova reverse shock. 

In this respect, the study of isolated SNRs evolving in diffuse low-temperature media has given insights 
on the dust mass fraction which is able to survive the passage of the reverse shock and make their way into the ISM 
(see, for example \citet{BianchiandSchneider2007,Nozawaetal2007,Marassietal2015,Micelottaetal2016,Bocchioetal2016} for 
theoretical estimates and \citet{Lauetal2015,GhavamianandWilliams2016,Lakicevicetal2015} for observational constraints). These 
studies agree that such fraction would lie between $0$ (completely destroyed) and $0.8$ depending on the ISM 
density and the mixing efficiency of the ejecta. 

\subsection{SNRs evolving in SSCs}

From the occurrence of a supernova explosion, the ISM is structured as follows: a region in 
which the ejecta expands freely; eventually two shocks are formed, a reverse shock (RS), which thermalizes the ejecta, 
and a leading shock (LS), which thermalizes the ISM; the shocked ejecta region is separated by a contact discontinuity 
from the matter swept-up by the leading shock. The thermalization of the kinetic energy is enhanced in young stellar 
clusters as stellar winds and supernova shells randomly collide and merge, thus creating a large central 
overpressure which results in the launching of a strong star cluster wind with temperatures $\gtrsim 10^6$ K and densities
$\sim (1-1000)$ cm$^{-3}$ \citep{ChevalierClegg1985,Silichetal2004,Silichetal2011,Wunschetal2011}. 

Thermal sputtering, promoted by the passage of the reverse shock, would lead to a size distribution favoring big grains 
($\gtrsim 0.05$ $\mu$m), as can be noted from the characteristic (graphite/silicate) grain lifetime 
against thermal sputtering in a hot gas, $\tau_{sput}$:

\begin{eqnarray}
      \label{eq:1}
\tau_{sput}  = \frac{a}{|\dot{a}|}=7\times 10^5 \frac{ a (\mu \text{m})}{n (\text{cm}^{-3})} 
\left[\left(\frac{T}{2\times 10^6 \text{ K}}\right)^{-2.5}+1 \right] , \nonumber \\
\end{eqnarray}

in units of years. In the above equation $a$ is the grain radius, $\dot{a}$ is the rate of decrease of the grain radius \citep{DraineandSalpeter1979,Tielensetal1994,TsaiMathews1995},
$n$ is the gas number density, and $T$ is the gas temperature. With ambient conditions $T=10^7$ K and $n=10$ cm$^{-3}$, 
a grain with radius $a=0.005$ $\mu$m would be completely eroded in $\sim 350$ years, while for $a=1$ $\mu$m, it would 
take $\sim 7.1\times 10^4$ years. However, shocks in a magnetized medium may also enhance the occurrence of frequent grain-grain 
collisions \citep{Jonesetal1996}. This can occur if the majority of grains reside in dense clumps and 
they move at high speeds relative to each other as a result of magnetohydrodynamic turbulence (see the discussion in Section 
\ref{sub:clumps}). This might imply that the population of small grains can be replenished from the shattering of long-living big 
grains. If this is the case, the post-shock size distribution will have an excess of small grains. Evidence 
in the $\sim$ 10$^4$ year-old galactic supernova remnant {\it Sgr A East} \citep{Lauetal2015} seems to point to an 
enhanced mass ratio between small grains and large grains (with characteristic sizes $\sim 0.001$ $\mu$m and $\sim 0.04$ $\mu$m, respectively), 
with values as high as $0.59$ and $0.90$ in the post-shock south clump and north regions.

As it was shown by \citet{MartinezGonzalezetal2016} (hereafter MTS16), the evolution of the grain size distribution 
promoted by thermal sputtering could greatly affect the appearance of the infrared spectral energy distributions (IR SEDs) 
of young stellar clusters, with the population of small grains playing a major role at NIR-MIR wavelengths. The 
importance of small grains ($\lesssim 0.05$ $\mu$m) on the emission properties of dusty media lies on their low heat capacities 
(as the heat capacity scales as $\sim a^3$). From the action of either electron-grain collisions (promoted in the hot 
and dense plasma streaming out as a star cluster wind) or the absorption of energetic photons (given their ample 
supply from a collection of hundreds or thousands of massive stars, \citealt{Takeuchietal2003}), small grains are more 
and more affected by stochastic temperature fluctuations compared to their bigger counterparts, provoking that they 
acquire temperatures from a few Kelvin up to $\sim10^3$ K, thus leaving their imprint at near-infrared (NIR) and 
mid-infrared (MIR) wavelengths. 

However, the conditions which allow collisional heating in a hot plasma, also promote thermal sputtering, with a greater
impact on small grains. \citet{Arendtetal2010} modeled the spectra of the Puppis A Supernova Remnant and showed that dust emission 
at short wavelengths ($\lambda \leq 20 \mu$m) is significantly reduced in the post-shock region (relative to the pre-shock region)
given an efficient small grain destruction.

NIR-MIR emission ($\sim 1-5$ $\mu$m), in excess to that of starlight and Polycyclic Aromatic Hydrocarbons (PAHs), has been 
observed in many nearby H\,{\sc ii} regions surrounding individual young stellar clusters, as in a number of the 
more than one hundred H\,{\sc ii} regions in M33 which properties have been thoroughly studied by \citep{Relanoetal2013,Relanoetal2016} 
(like NGC 604 and NGC 595, to name the most prominent examples) and in individual clusters in SBS 0335-052E \citep{Reinesetal2008}. 
In fact, \citet{Relanoetal2016} and \citet{Reinesetal2008} required an additional $\sim 1000$-K blackbody component of 
``uncertain origin'' to fit their observed IR SEDs.

Similar excesses, more prevalent in the MIR, have also been observed in a number of blue compact dwarf galaxies (e.g: I Zw 18, II Zw 40, Mrk 930 and Haro 11 
\citep{RemyRuyeretal2015,Vanzietal2000,Izotovetal2014}). Therefore, these observations not only point to the existence of a very hot 
dust component ($\sim$400-1000 K), but also give compelling evidence for the presence of a persistent 
(i.e. long-living and/or constantly replenished) population of small grains associated to young stellar clusters. 

From these motivations, here we advocate that a population of SN-condensed small grains, heated by the intense 
radiation field emerging from the cluster and frequent electronic collisions in the thermalized ejecta, is 
responsible for the observed NIR-MIR excesses in young massive clusters, in particular in the case of 
the young massive clusters in M33. For this purpose, we have considered a number of physical effects which 
might prolong the emergence of the NIR-MIR excess, e.g. a quick replenishment and mitigated destruction of the population of small 
grains.

The paper is organized as follows: in Section \ref{sec:hot_dust} we introduce of our star cluster and star cluster wind models 
(subsection \ref{sub:WInd}), the consideration of grain processing in clumpy SN ejecta (\ref{sub:clumps}), the formulation of 
the supernova dust injection process and its relevant timescales are presented 
(subsection \ref{sub:injection}). In Section \ref{sec:M33} we give the reasons to apply our model to the infrared spectral 
energy distributions to the most prominent H\,{\sc ii} regions in M33 and review some of their relevant properties (stellar 
mass, spectral indexes, age) obtained from the literature; we also briefly discuss the treatment of the 
additional dust components, other than the newly-produced grains, required to fit the observed IR SEDs in the whole 
infrared regime (subsection \ref{sub:HII}). Section \ref{sec:results} deals with the results of our models and the requirements of the model to work. 
Our results are presented at different evolutionary times. In Section we discuss the relative importance of
other dust sources compared to SN dust. The summary of the main ingredients of the model
are presented in Section \ref{sec:Summary}, while our conclusions appear in Section \ref{sec:conclusions}. 

\section{Hot Dust within the Star Cluster}
\label{sec:hot_dust}
\subsection{Star Cluster Wind Model}
\label{sub:WInd}

We focus on coeval young massive stellar clusters with ages 4-6 Myr, i.e, at the start of their supernova era
\citep[with the first SN occurring at $\sim 3.5$ Myr, e.g. ][]{Krauseetal2013, MartinezGonzalezetal2014, Wunschetal2017}. At
$\sim 6$ Myr, all stars with masses $\geq 40$ M$_{\odot}$ should have exploded as supernovae \citep{MeynetMaeder2003}. For simplicity, 
the SN explosions are considered to occur a the very center of spherically-symmetric young star clusters with a 
Schuster stellar density distribution of the form \mbox{$\rho_* \propto  [1+(r/R_{c})^2]^{-\beta}$}, where $\beta$ is taken 
to be $1.5$, $r$ is the distance from the cluster center, $R_{c}$ is the core radius and $R_{SC}$ is the star cluster 
truncation radius. We calculate the gas number density and temperature inside the star 
cluster by making use of the model thoroughly discussed in \citet[][]{Silichetal2011} and 
\citet[][]{Palousetal2013} which solves the set of hydrodynamic equations for the stationary flow driven 
by stellar winds and supernova explosions. Our models include the effects of gas \citep{Raymondetal1976} and dust radiative 
cooling \citep{Dwek1987,TenorioTagleetal2013,TenorioTagleetal2015,MartinezGonzalezetal2016}. The star cluster mechanical luminosity, $L_{SC}$, 
and the stellar mass of the cluster, $M_{SC}$, are related as $L_{SC}= 3 \times 10^{39} (M_{SC}/10^5 \mbox{M}_\odot)$ erg s$^{-1}$ 
\citep{Leithereretal1999}. The mechanical energy is treated as a constant due to the short timescales, on 
the order of thousands of years, investigated in this work. Within the cluster, mass is reinserted via stellar winds
and SN explosions at a rate $\dot{M} = 2 L_{SC}/V_{A\infty}^2$, where $V_{A\infty}$ is the adiabatic wind terminal speed. 

\begin{figure}
\begin{center}
\includegraphics[scale=0.7]{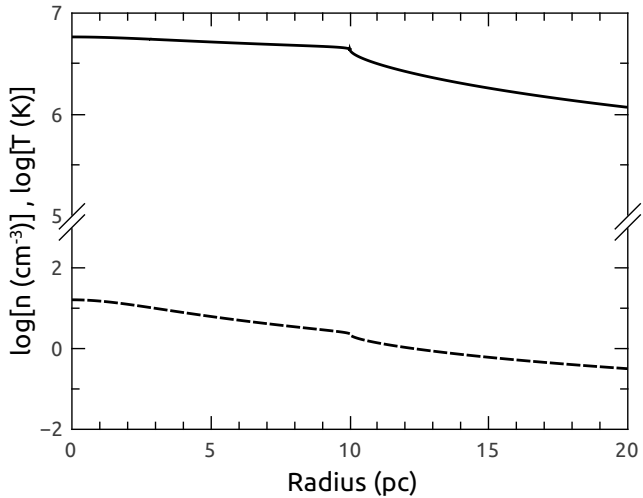}
\caption{The radial distribution of temperature and density in the star cluster as calculated from our hydrodynamical model
\citep{Silichetal2011,Palousetal2013}.The upper and lower curves correspond to the calculated temperature and density, respectively,
for a cluster with $L_{SC}=1.7\times 10^{40}$ erg s$^{-1}$, $R_{c}=3$ pc, $R_{SC}=10$ pc and $V_{\eta \infty}= 630$ km s$^{-1}$. 
This model is later applied to the case of the central cluster in NGC 604 (see Section \ref{sec:M33}). Note the axis break
and the change of units in the upper and lower regions of the plot.}
\label{fig:prof}
\end{center}
\end{figure}

One can also consider two important parameters: heating efficiency ($\eta_{he}$) and mass loading ($\eta_{ml}$). The 
former represents the fraction of $L_{SC}$ which is converted into the thermal energy of the wind \citep[see e.g.][]{Silichetal2007}. 
The latter arises under the condition of a non-negligible mass of gas left over from star formation which is incorporated into the 
star cluster wind at a rate $\dot{M}_{inj}=(1+\eta_{ml})\dot{M}$ \citep[see e.g.][]{Wunschetal2011}. 
After the consideration of $\eta_{he}$ and $\eta_{ml}$, the value of the wind terminal speed changes to
$V_{\eta \infty}=V_{A \infty}[\eta_{he}/(1+\eta_{ml})]^{1/2}$.

In Figure \ref{fig:prof}, we show the radial distribution of temperature and density of the wind in a star cluster with
$L_{SC}=1.7\times 10^{40}$ erg s$^{-1}$, $R_{c}$ and $R_{SC}$ equal to $3$ and $10$ pc respectively, and $V_{\eta \infty}= 630$ km s$^{-1}$. 
This model is later applied to the case of the central cluster in NGC 604 (see Section \ref{sec:M33}).

Given a low heating efficiency, grain thermal sputtering is reduced as a consequence of a lower temperature in the 
shocked wind, alleviating the destruction of the population of small grains necessary to explain the 
NIR-MIR excesses. In favor of a low heating efficiency, first proposed by \citet{Smithetal2006} and \citet{Larsenetal2006}, one can 
mention the work of \citet{Silichetal2007,Silichetal2009}, who studied several young clusters in the central zone of M82. 
They found that the observed sizes and expansion velocities of the associated H\,{\sc ii} regions are not consistent 
with the standard bubble model \citep{Weaveretal1977} which might be the result of a low heating efficiency ($\lesssim 0.1$). 

\subsection{Grain Processing in Clumpy Ejecta}
\label{sub:clumps}
In star clusters, SN-condensed grains residing in dense clumps might be efficiently 
accelerated by the magnetohydrodynamic (MHD) turbulence sustained by massive stars as shown by \citet{Hirashitaetal2010}. For instance, they 
found that a grain (silicate or carbonaceous) with radius $0.1$ $\mu$m in an ionized warm medium with density between $1$ and $10$ cm$^{-3}$
moves at a speed of a few km s$^{-1}$ with respect to another grain with radius $0.01$ $\mu$m (see their Figure 1). Moreover, 
as the clumps in the ejecta are traversed by the reverse shock, the grains would experience a further size-dependent acceleration 
as they gyrate along the compressed magnetic field lines \citep{Yanetal2004,Shull1977}.

On this basis, one can anticipate the relevance of grain-grain collisions and the modification of the grain size distribution in clumpy SN 
within star clusters. If we assume that all the newly-condensed grains reside in clumps with 
radius $500$ AU and a mass of dust per clump $M_{dcl} = 10^{-6}$ M$_{\odot}$ \citep[average values for the dusty globules in the Crab 
Nebula, ][]{Grenmanetal2017}. Then, the average dust number density inside each clump is

\begin{eqnarray}
\langle n_d \rangle = \frac{3 M_{dcl}}{\displaystyle 4 \pi \langle m_{d} \rangle R_{cl}^3} ,
\end{eqnarray}

where $\langle m_{d} \rangle$ is the mean mass per grain. For a size distribution of the grains in the unshocked clumps 
characterized by a power-law, $\sim a^{-\alpha}$ with $\alpha=3.8$ 
\citep[][\footnote{\citet{TemimDwek2013} found that the grain size distribution for the Crab Nebula is characterized by a power-law 
index between -3.5 and -4.0}]{TemimDwek2013} and lower and upper limits $a_{min}=0.001$ $\mu$m and $a_{max}=0.1$ $\mu$m, respectively,
$\langle m_{d} \rangle$ is $\approx 2.9\times 10^{-19}$ g (for a grain density equal to $3.3$ g cm$^{-3}$), thus the average grain 
size is $\langle a \rangle \approx 0.0027$ $\mu$m and $\langle n_d \rangle \approx 3.9\times 10^{-3}$ cm$^{-3}$. 

The timescale for an encounter between a grain with radius $a$, and an average grain of radius $\langle a \rangle$ in a 
clump is 

\begin{eqnarray}
\Delta t_{coll}=\left[ \pi (a+\langle a \rangle)^2 \langle n_d \rangle \delta v \right]^{-1},
\end{eqnarray}

where $\delta v$ is the mean relative velocity between grains. If $\delta v$ is higher than $2.7$ km s$^{-1}$ 
for silicate and $1.2$ km s$^{-1}$ for graphite grains \citep{Jonesetal1996}, as it is in the case of the 
MHD turbulence predicted by \citet{Hirashitaetal2010}, then the grain population is effectively influenced by 
shattering during the lifetime of the clumps. For instance, if $\delta v \approx 5$ km s$^{-1}$, a 
silicate grain with radius $0.1$ $\mu$m will encounter another grain every 14 days. 

The lifetime of individual ejecta clumps is largely defined by the propagation of the reverse shock across them, 
leading to their fragmentation and rapid destruction. The dynamical time for this process has been estimated by 
\citet{Micelottaetal2016}. For a clump with radius $R_{cl}\sim 500$ AU, a density contrast between the clump and the 
smooth ejecta $\sim 100$, and the (attenuated) propagation velocity of the shock within the clump 
$V_{cl} \approx 150$ km s$^{-1}$, the clump would be destroyed in $t_{dest} = 3.5 R_{cl}/V_{cl} \approx 55$ years. This is 
a very short time compared to the crossing time of the reverse shock through the whole ejecta ($\lesssim 3\%$ in all our 
calculations, see Section \ref{sub:injection}), but also large compared to $\Delta t_{coll}$. The implication is that by the 
time when thermal sputtering commences to act in the thermalized ejecta, the grain size distribution has already been shifted towards smaller grains.
One can then assume that grains with radius $\geq 0.05$ $\mu$m have all been shattered,
thus producing an excess of grains with radius $\leq 0.001$ $\mu$m. From the above considerations, in our 
calculations we have taken the grain size distribution immediately after clump destruction to have the form 
$\sim a^{-\alpha}$ ($\alpha=3.8$) with lower and upper limits $a_{min}=0.0005$ $\mu$m and $a_{max}=0.05$ $\mu$m.

In characterizing the emission from grains residing in unshocked clumps, we consider that silicate and graphite grains
(assumed to be produced in equal proportions) are only heated by the absorption of the radiation field emerging from 
the cluster, whereas grains in the thermalized medium (see next Section) are also subject to heating induced by 
collisions with free electrons.

\subsection{Dust Injection into the Thermalized Medium}
\label{sub:injection}

We follow the evolution of the graphite and silicate grain populations under the action of thermal 
sputtering in a similar manner (albeit more physically motivated) as \citetalias{MartinezGonzalezetal2016}. 
Once the gas from the ejecta is thermalized, its temperature and density is similar to that of 
the shocked star cluster wind, and thus they are considered to be the same medium. In the subsequent, we will 
refer to the SN ejecta and the cluster wind simply as the thermalized medium.

Dust injection into the thermalized medium starts once individual clumps in the ejecta are destroyed,
and finishes when all the ejecta is thermalized (at $t=\tau_{inj}$). This occurs in a 
timescale (in units of years) given by \citet{TangandWang2009} 
\citep[see also][]{ReynoldsandChevalier1984} for a supernova remnant evolving 
in a hot medium:

\begin{equation}
\label{eq:tinj}
  \tau_{inj} \simeq 10^{4} \left(\frac{\rho}{\rho_s}\right)^{-1/3}  \left(\frac{M_{ej}}{1.4 \text{ M}_\odot}\right)^{5/6}
  \left(\frac{E_{SN}}{10^{51} \text{ erg}}\right)^{-1/2}  ,
\end{equation}

where $\rho=1.4 m_{H} n$ is the gas density, $m_{H}$ is the H mass, $\rho_s=1.67\times 10^{-26}$ g cm$^{-3}$, 
$M_{ej}$ is the ejecta mass and $E_{SN}$ is the kinetic energy of the ejecta. At $\tau_{inj}$, the radius
of the leading shock can be expressed as:

\begin{equation}
\label{eq:rls}
  R_{LS}(\tau_{inj}) = \xi \left(\frac{E_{SN}\tau_{inj}^2}{\rho} \right)^{1/5} 
  {}_{2}F_{1}\left(\frac{-3}{5},\frac{2}{5};\frac{7}{5};-\frac{\tau_{inj}}{t_{c}}\right) , 
\end{equation}

where ${}_{2}F_{1}$ is the Gauss hypergeometric function which accounts for a non-negligible ambient 
pressure \citep{TangandWang2005}, $\xi$ is equal to $1.15$ (for an ideal gas with a ratio of specific
heats $\gamma=5/3$) and $t_{c}$ is defined as:

\begin{equation}
\label{eq:tc}
  t_{c} = \left[ \left(\frac{2}{5}\xi \right)^5 \frac{E_{SN}}{\rho c_{s}^5}  \right]^{1/3} , 
\end{equation}

with $c_{s}$ representing the ambient medium sound speed. For reference, if $n= 5$ cm$^{-3}$,
$T=5.5 \times 10^6$ K, $M_{ej}=8$ M$_\odot$ and $E_{SN}=$10$^{51}$ erg, the injection timescale would be 
$\tau_{inj}\sim 3820$ years when the leading shock would have reached $R_{LS}\sim4.90$ pc. 

\begin{figure*}
\begin{center}
\includegraphics[scale=0.45]{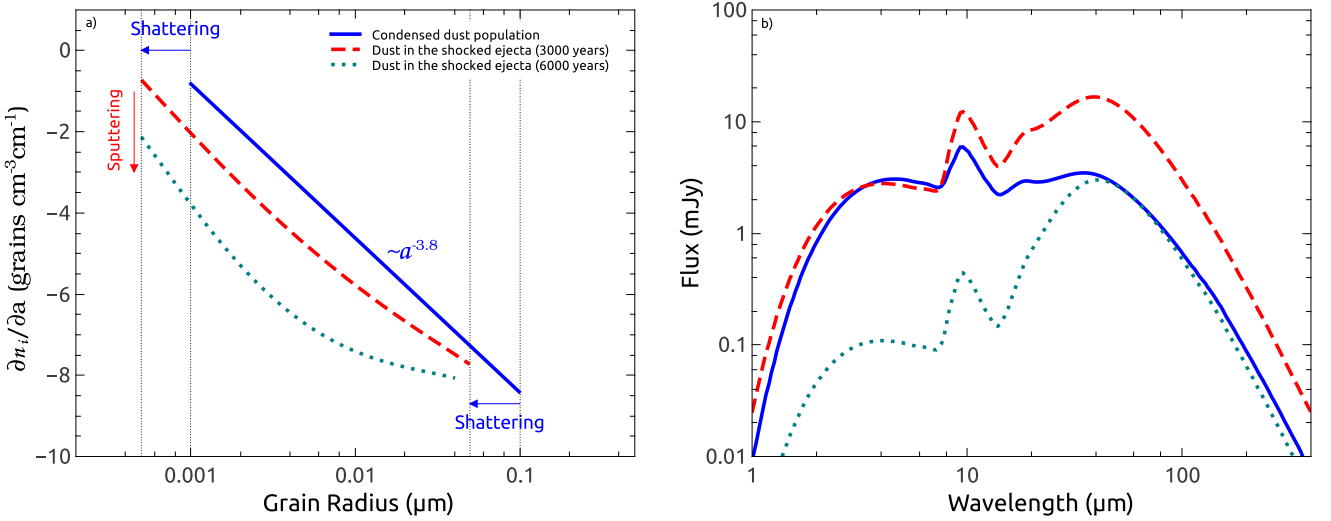}
\caption{The Evolution of the Grain Size Distribution and the associated IR emission. Panel a: The blue solid line displays the size distribution of the condensed dust population.  The red-dashed (green-dotted) line corresponds to the size distribution associated to dust grains already incorporated into 
the thermalized medium at 3000 (6000) years after the dust injection started. Grain shattering shifts the distribution towards smaller grains 
($a_{min}=0.0005$ $\mu$m and $a_{max}=0.05$ $\mu$m, indicated by blue arrows) while thermal sputtering lowers it (indicated by a red arrow) and changes 
its slope. Grains smaller than (bigger than) $\sim 0.005$ $\mu$m are less (more) affected by ionic collisions and thus the size distribution is not simply
described by a power-law. The upper and lower limits of the size distributions are marked with dotted vertical lines. Panel 
b shows the IR emission associated to the grain size distributions presented in panel a (see the text for the description of the input model 
parameters).}
\label{fig:size}
\end{center}
\end{figure*}

From the consideration of dust being incorporated into the thermalized medium and the action of thermal 
sputtering, the grain size distribution, $\displaystyle \partial n_{i}/\partial a$ (where $n_{i}$ is the 
grain number density of each dust species), evolves with time from the continuity equation 
(\citet*{LaorDraine1993};\citet*{YamadaKitayama2005}:

\begin{eqnarray}
\label{eq:2}
\dot{a}\derp{}{a} \left(\derp{n_{i}}{a} \right) +
\derp{}{t} \left(\derp{n_{i}}{a} \right) 
=  A_{i}^{(m)} a^{-\alpha}/\tau_{inj}^{(m)} , 
\end{eqnarray}

if $t\leq \tau_{SN}^{(m)}+\tau_{inj}^{(m)}$, where $\alpha$ is the index of the dust size distributions 
immediately after injection into the thermalized medium, $\tau_{inj}^{(m)}$ is the injection timescale for 
the $m$-supernova event which occurs at $t=\tau_{SN}^{(m)}$. The normalization constants, $A_{i}^{(m)}$ (cm$^{\alpha-4}$), 
are given by:

\begin{eqnarray}
      \label{eq:2c}
A_{i}^{(m)} = \frac{\displaystyle f_{i} M_{dSN}^{(m)}/V_{SC}}{\displaystyle \int^{a_{max}}_{a_{min}} 
     \frac{4}{3} \pi \rho_{gr} a^{3-\alpha} \mbox{ d}a}   ,
\end{eqnarray}

where $\rho_{gr}$ is the grain density, $f_{i}$ is the species mass fraction, $M_{dSN}^{(m)}$ is the total 
mass of dust formed/injected in a single supernova and $V_{SC}$ is the star cluster volume. The constants 
$A_{i}^{(m)}$, are zero until $t=\tau_{SN}^{(m)}$.

If the reverse shock has already reached all 
the dust produced by the $m$-supernova $(t>\tau_{SN}^{(m)}+\tau_{inj}^{(m)})$, i.e. dust injection has 
ceased, the right-hand term in equation \eqref{eq:2} is set to zero. General solutions of equation \eqref{eq:2} 
corresponding to $j$ dust injections are given by \citepalias{MartinezGonzalezetal2016}:

\begin{eqnarray}
\label{eq:2a}
\derp{n_{i}}{a}
= \displaystyle \sum\limits_{m=1}^j \frac{A_{i}^{(m)}}{\tau_{inj}^{(m)}\dot{a}} \Biggr\{&& \frac{a^{-\alpha+1}}{(-\alpha+1)} 
        \nonumber \\  &&    - \frac{\left[a-\dot{a}(t-\tau_{SN}^{(m)})\right]^{-\alpha+1}}{(-\alpha+1)} \Biggr\} , 
\end{eqnarray}

for $t\leq \tau_{SN}^{(m)}+\tau_{inj}^{(m)}$, and after the $m$-dust injection (for 
$t>\tau_{SN}^{(m)}+\tau_{inj}^{(m)}$):

\begin{eqnarray}
\label{eq:2b}
\derp{n_{i}}{a}
= \displaystyle \sum\limits_{m=1}^j  \frac{A_{i}^{(m)}}{\tau_{inj}^{(m)}\dot{a}} \Biggl\{ && \frac{\left[a-\dot{a}(t-\tau_{SN}^{(m)}-\tau_{inj}^{(m)})\right]^{-\alpha+1}}{(-\alpha+1)}
  \nonumber \\ &-& \frac{\left[a-\dot{a}(t-\tau_{SN}^{(m)})\right]^{-\alpha+1}}{(-\alpha+1)} \Biggr\} .
\end{eqnarray} 

The interstellar radiation field emerging from the star cluster, $J_{\lambda}$, is characterized by multiples,
$U$, of the solar neighborhood radiation field, $J_{\lambda}^{NBH}$ \citep{Mathisetal1983}. The infrared flux per unit 
wavelength, produced by a population of dust grains with the same chemical composition, from a source located at distance 
$D_{SC}$, is given by:

\begin{eqnarray}
      \label{flambda}
       \hspace{-1.1cm}
f_{\nu} &=& \left(\frac{1.4 m_{H} Z_{d} N_{H}}{\rho_{d}} \right) \pi \Omega_{SC} \times \nonumber \\ 
           && \int_{a_{min}}^{a_{max}}\int_{0}^{\infty} a^{2} \derp{n_{i}}{a} Q_{\nu}(a) B_{\nu}(T_d) G(a,T_d) \mbox{ d}T_{d} \mbox{ d}a , \nonumber \\
\end{eqnarray}

in units erg s$^{-1}$ cm$^{-2}$ Hz$^{-1}$ \citep{DwekandArendt1992}. In the above equation, $N_{H}$ is 
the hydrogen column density, $\rho_{d}$ is the size-averaged grain density, $\Omega_{SC}$ is the solid angle subtended by the cluster, 
$T_{d}$ is the dust temperature, $G(a,T_d)$ is the dust temperature distribution resulting from stochastic temperature 
fluctuations produced by both, photon absorptions and electron-grain collisions 
\citep[see ][]{GuhathakurtaandDraine1989,Dwek1986}, $Q_{\nu}(a)$ is the dust absorption efficiency 
and $B_{\nu}$ is the Planck function. Additionally, $Z_{d}$ is the time-dependent dust-to-gas mass ratio 
\citepalias{MartinezGonzalezetal2016}. 

After the dusty clumps are reached by the RS, frequent grain-grain collisions induced by the high grain density inside the clump and the relative 
motion of the grains in a turbulent ejecta, lead to the very quick fragmentation of the large grains into smaller pieces, thus shifting the size 
distribution towards smaller grains. Once the clumps are disrupted, shattering is no longer efficient and thermal sputtering dominates the evolution 
of the grain size distribution according to equations \eqref{eq:2a} and \eqref{eq:2b}.

However, we have made use of the size-dependent correction to the sputtering yield 
(the number of atoms removed from a grain after a collision with an energetic particle) as prescribed by \citet{SerraDiazCanoetal2008}
(approximated by the formula given by \citealt{Bocchioetal2012} and averaged by the Maxwell-Boltzmann energy distribution) 
in order to consider more realistic sputtering rates. This correction has the ability of increasing the derived sputtering 
yields for medium-sized and big dust grains, while reducing it for very small grains as they become transparent to the more 
energetic incident ions, thus mitigating their destruction. The latter effect results as the implantation depth of the 
incident ion (the characteristic radius at which the ion is stopped), becomes larger than the grain diameter 
(for sufficiently energetic particles, see equation A.2 in \citealt{Bocchioetal2012}). Hence, the resultant grain size distribution
is no longer described by a simple power-law.

Figure \ref{fig:size} (panel a) presents the evolution of the size distribution of a population of $0.1$ M$_{\odot}$ of dust injected into the 
intracluster medium. Initially, the size distribution (just after grain condensation) is characterized by a power-law $\sim a^{-\alpha}$ with 
$\alpha=3.8$, $a_{min}=0.001$ $\mu$m and $a_{max}=0.1$ $\mu$m. The grain population in this case is heated by a radiation field $2\times 10^5$ 
times that of the solar neighborhood value \citep{Mathisetal1983}. Panel a also shows the size distributions corresponding to grains 
already injected into the thermalized medium, with number density $2.5$ cm$^{-3}$ and temperature $2.6\times 10^6$ K, at $t=3000$ and $6000$ 
years. 

Panel b in Figure \ref{fig:size} displays the IR emission associated to the grain population depicted in panel a. There is a strong NIR-MIR 
emission from dust heated by starlight in the pre-shocked ejecta which can account for the observed NIR-MIR excess. However, their emission 
will be rapidly overcome by that of the dust in the post-shock region. Given that the small grains are rapidly destroyed by thermal sputtering, 
the NIR part of the emission will be greatly reduced after a few thousand years.

As mentioned before, we restrict our models to the case of supernovae occurring at (or very close to) the center  
of the star cluster. In the case when this assumption is relaxed, off-centered supernova blast waves would 
be evolving in the steep density gradient expected at large radii in the cluster, leading to the blowout and rapid loss
of energy of the ejecta \citep{TenorioTagleetal2015b,Silichetal2017} and the release of the SN products out 
of the stellar cluster, possibly including a non-negligible amount of dust grains. Finally, although thermal sputtering may produce grains 
with radii $\leq 0.0005$ $\mu$m, these grains are removed from all our calculations as they are considered completely sputtered.

\section{The Star Clusters in M33}
\label{sec:M33}

With the considerations expressed in the prior sections, we have applied our model to the observed 
IR SEDs of the M33 clusters (located at a distance $\sim 817\pm 58$ kpc \citealt{Freedman2001}), in particular to 
the prominent regions NGC 604, NGC 595, NGC 588, and NGC 592 (regions 98, 44, 7, 25 in the \citet{Relanoetal2016} 
sample, respectively). Our intention is not to provide best fitting models to the observed IR SEDs; rather, we aim to 
show that a population of newly-produced supernovae grains, subject to the extreme prevailing conditions in young 
massive clusters, is able to explain the NIR-MIR excesses observed in such objects. We selected these four regions, because 
they all have reliable estimates of their stellar masses \citep{Relanoetal2010} and have experienced recent supernova activity 
as can be inferred from their relatively flat spectral indexes 
\citep[][see Table \ref{tab:1}]{Tabatabaei2007,Yangetal1996,Gordonetal1993}, which is interpreted as non-thermal 
(synchrotron) emission from supernova remnants. 

In Table \ref{tab:1} we present the stellar masses for the four regions, their spectral indexes (denoted as $\psi$), 
the total number of stars in the respective cluster, the number of massive stars with masses $\geq 40$ M$_{\odot}$,
the average interval between successive supernova explosions ($\langle \Delta \tau_{SN}\rangle$, the inverse of the average 
supernova rate), as predicted by the Geneva evolutionary tracks \citep{Meynetetal1994} implemented in the Starburst99 synthesis 
model \citep{Leithereretal1999} for a 4-Myr star cluster and a \citetalias{Kroupa2001} initial mass function 
(indexes $-1.3$ and $-2.3$) with lower and upper cut-off mass of 0.1 M$_{\odot}$ and 100 M$_{\odot}$, respectively, a turn-off 
mass at 0.5 M$_{\odot}$ and metallicity Z=0.4Z$_{\odot}$.

\begin{deluxetable}{lccccc}
\tablecolumns{6} \tablewidth{0pc}
\tablecaption{\label{tab:1} \sc Cluster Properties}
\tablehead{
\colhead{Region} & \colhead{$M_{SC}$} & \colhead{$\psi$} & \colhead{No. of Stars} & \colhead{No. of Stars} &\colhead{$\langle \Delta \tau_{SN}\rangle$} \\
\colhead{}  & \colhead{{\tiny ($10^4$M$_{\odot}$)}} & \colhead{-} & \colhead{-} & \colhead{\tiny($\geq 40$M$_{\odot}$)} & \colhead{years}
}
\startdata
NGC 604 & $56.8$& 0.12 & 6.5$\times10^5$ & 650 &3800    \\
NGC 595 & $22.4$& 0.07 & 2.6$\times10^5$ & 250 &8250    \\
NGC 588 & $6.86$& 0.00 & 7.9$\times10^4$ & 80 &31000   \\
NGC 592 & $3.98$& 0.13 & 4.6$\times10^4$ & 45 &50000 
\enddata
\tablecomments{The Table presents the stellar masses of the selected star clusters 
obtained by \citet{Relanoetal2010}, spectral indexes as estimated by \citet{Tabatabaei2007}. For the average interval between supernova explosions 
we ran a Starburst99 model with a standard \citetalias{Kroupa2001} Initial Mass Function with metallicity Z=0.4Z$_{\odot}$, and Geneva evolutionary tracks \citep{Meynetetal1994} 
with no rotation.}
\end{deluxetable}

From the consideration of star clusters with masses $\lesssim 10^5$ M$_{\odot}$, the obtained dust injection timescales
represent only a minor fraction of the interval between supernova explosions, making the replenishment of grains difficult.
This is why we expect that some important fraction of these clusters (with masses similar to NGC 588 and 
NGC 592 or lower) should present a marginal or non-existent NIR-MIR excess with respect to the emission from PAHs.
In such cases another problem arises; the typical mass of a SN ejecta (in our case assumed to be 
$5$ M$_{\odot}$, see e.g. \citealt{Yadavetal2016}) becomes comparable to the gas mass enclosed in the star 
cluster volume. If this is the case, the capture of the SN ejecta is not achieved and a smooth cluster wind is not 
developed; rather the supernova explosions behave as isolated events \citep{Sharmaetal2014}
where the dust content is diluted in a larger volume and thermalization is inefficient. To overcome this problem, it is 
necessary not only to consider a low heating efficiency to reduce the effects of thermal sputtering, but also the effect 
of mass loading in the stellar wind to warrant that mass enclosed inside $R_{SC}$ is always significantly larger than 
the mass ejected by a single supernova. For this reason, our models for NGC 588 and NGC 592 include mass loading, however,
in conservative values ($\eta_{ml}=1$).

The hydrodynamical model for NCG 604, consists of a star cluster with a mechanical 
luminosity $\sim1.71 \times 10^{40}$ erg s$^{-1}$, with dimensions given by $R_{c}=3$ pc and $R_{SC}=10$ pc 
(which give a half-mass radius $R_{HM}=5.4$ pc), an average interval between successive supernova 
explosions $\sim 3800$ years, an adiabatic wind terminal speed $V_{A \infty}=2000$ km s$^{-1}$ (which with 
$\eta_{he}=0.1$ is reduced to $V_{\eta \infty}=632.5$ km s$^{-1}$), lower and upper limits for the injected 
dust size distribution, $a_{min}= 0.0005 \mu$m and $a_{max}= 0.05 \mu$m, respectively. The average conditions, 
gas density and temperature, for the thermalized medium  in this case are $\langle n \rangle \sim 7.4$ cm$^{-3}$ 
and $\langle T \rangle =5.2\times 10^6$ K. For typical supernova values ($M_{ej}=8$ M$_{\odot}$, $E_{SN}=10^{51}$ 
erg s$^{-1}$) and the prevailing conditions inside the cluster, evaluation of equation \eqref{eq:tinj} results in 
an injection timescale $\tau_{inj}=2800$ years, when the SN leading shock radius is about $4$ pc and is 
well-contained inside $R_{SC}$. The main input parameters used in 
our four models are summarized in Table \ref{tab:2} and the derived quantities from the hydrodynamical model 
are presented in Table \ref{tab:3}. 

\begin{deluxetable}{lcccc}
\tablecolumns{5} \tablewidth{0pc}
\tablecaption{\label{tab:2} \sc Input Parameters}
\tablehead{
\colhead{Region} & \colhead{$L_{SC}$} & \colhead{$\eta_{he}$} & \colhead{$\eta_{ml}$} & \colhead{$U$} \\
\colhead{}  & \colhead{{\tiny($10^{39}$erg s$^{-1}$)}} & \colhead{-} & \colhead{-} & \colhead{-} 
}
\startdata
NGC 604 &  $17.1$ &    $0.1$   &    $0$    & $1\times10^6$  \\
NGC 595 &  $6.72$ &    $0.1$   &    $0$    & $5\times10^5$  \\
NGC 588 &  $2.05$ &    $0.1$   &    $1$    & $2\times10^5$  \\
NGC 592 &  $1.19$ &    $0.1$   &    $1$    & $1\times10^5$
\enddata
\tablecomments{The Table presents the input parameters we used for the selected clusters. The values of $L_{SC}$
were obtained from the relation $L_{SC}= 3 \times 10^{39} (M_{SC}/10^5 \mbox{M}_\odot)$ erg s$^{-1}$ 
\citep{Leithereretal1999}. $U$ characterizes the strength of the radiation field in multiples of the solar neighborhood value, 
$J_{\lambda}^{NBH}$ \citep{Mathisetal1983}.}
\end{deluxetable}

For NGC 604 and NGC 595, which stellar content is above $10^5$ M$_{\odot}$ and the estimated age is $\sim 4$ Myr 
\citep{Relanoetal2013}, the radiation field emerging from them is assumed to be $10^6$ and $5\times 10^5$ times that of 
the solar neighborhood, while for regions with stellar masses lying below $10^5$ M$_{\odot}$, the radiation field is 
assumed to be $2\times10^5$ and $10^5$ times the solar neighborhood radiation field, respectively. In all our models,  
dust emission in the thermalized medium is dominated by collisional heating at NIR wavelengths, while photon heating at 
MIR wavelengths. We also note that in a coeval cluster, the number of UV photons starts to drop (as $\sim t^{-5}$, 
\citealt{Beltramettietal1982}) as a consequence of the explosion of the massive stars \citep{MartinezGonzalezetal2014}. 

For simplicity, all of the modeled clusters are assumed to be contained in the same volume. The average density and 
temperature inside the cluster in each case are shown in Table \ref{tab:3}.

\begin{deluxetable}{lcccc}
\tablecolumns{5} \tablewidth{0pc}
\tablecaption{\label{tab:3} \sc Model Outputs}
\tablehead{
\colhead{Region} & \colhead{$V_{\eta \infty}$} & \colhead{$\langle n \rangle$} & \colhead{$\langle T \rangle$} & \colhead{$\tau_{inj}^{(m)} / \langle \Delta \tau_{SN}\rangle$} \\
\colhead{}  & \colhead{{\tiny(km s$^{-1}$)}} & \colhead{{\tiny (cm$^{-3}$)}} & \colhead{{\tiny ($10^6$ K)}} & \colhead{-} 
}
\startdata
NGC 604 & $630$ &   $7.4$   & $5.2$    & 0.50   \\
NGC 595 & $630$ &   $2.9$   & $5.3$    & 0.50   \\
NGC 588 & $450$ &   $2.5$   & $2.6$    & 0.15   \\
NGC 592 & $450$ &   $1.5$   & $2.6$    & 0.10   
\enddata
\tablecomments{The values of $V_{\eta \infty}$ were obtained from an $V_{A \infty}=2000$ km s$^{-1}$ and the corresponding 
assumed values of $\eta_{he}$ and $\eta_{ml}$ presented in Table \ref{tab:2}. The average gas densities and temperatures were 
obtained from the hydrodynamical star cluster wind model. Finally, we present the ratio of the dust injection timescale to the 
average interval between successive supernova explosions in the star
cluster.}
\end{deluxetable}

\subsection{The Cold, Warm Dust and PAH\textsc{s} Components}
\label{sub:HII}

Following \citet{Relanoetal2016}, we have used three additional components to account for the observed 
infrared spectral energy distributions. These components arise from the H\,{\sc ii} regions surrounding the central
star clusters. One component is that from cold dust radiating at low temperatures (a few times $\sim 10 $ K). From this 
component a huge amount of dust can be inferred (thousands or tens of thousands of solar 
masses), most probably the result of efficient grain growth \citep[see for example][]{Asanoetal2013}. A second component corresponds to
warm dust ($\sim (50-100)$ K) which \citet{Relanoetal2016} attributes to the emission of very small grains. The third component is that
from Polycyclic Aromatic Hydrocarbons (PAHs). The heating of this grains is always dominated by the absorption of the radiation 
field, while collisional heating is always negligible due the low temperature of the diffuse medium in which the grains
are immersed. For simplicity, in fitting the observed data in the case of the warm and cold components, we have followed the 
approach taken by \citet{Mattssonetal2015}. In their approach, one assumes a dust temperature distribution 
instead of a dust size distribution (for the hot dust component we did the opposite). With the method, 
one does not have to assume a certain radiation field to derive a dust mass. Our chosen dust temperature 
distribution is represented by a power-law of the form $G(a,T_{d}) \propto T_{d}^{-3-\chi/2}$, where $\chi$ is 
the effective emissivity index which we take as 1. 

For the cold and warm dust components, we have set the limits of the size distribution to $0.0001$ $\mu$m and 
$0.5$ $\mu$m for all the models. The temperature range for our NGC 604 and NGC 595 cold dust models was set 
to $(16-33)$ K, while for the NGC 588 and NGC 592 models it was set to $(12-33)$ K.
In the case of the warm dust component the range of temperatures is ($50-100$) K for all the models. Finally, 
for the PAHs component, the range of sizes was taken from $0.0001$ $\mu$m to $0.0009$ $\mu$m while the range of 
temperatures was set to ($90-750$) K in all the models.

\section{Results}
\label{sec:results}

\begin{figure}
\begin{center}
\includegraphics[scale=0.7]{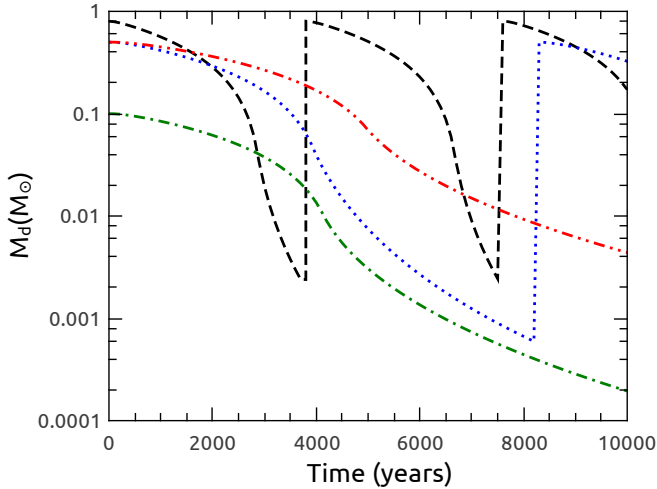}
\caption{The evolution of the dust mass in the star cluster. The curves display the total mass of dust 
present in the star cluster as a function of time for our NGC 604 (black dashed line), NGC 595 
(blue dotted line), NGC 588 (green dash-dotted line) and NGC 592 (red dash-double dotted line) models. 
In the models, the mass of dust injected inside the star cluster is $0.8$, $0.5$, $0.1$ and $0.5$ M$_{\odot}$,
respectively. Note that almost all the dust mass is returned to the gas phase in a timescale of a few thousand 
years as a result of efficient thermal sputtering.}
\label{fig:Md}
\end{center}
\end{figure}

In Figure \ref{fig:Md}, we show the evolution of the total mass of dust within the central star 
cluster ($\leq R_{SC}$) as a result of successive supernova explosions. The mass of dust produced in the ejecta of 
individual supernovae is taken to be $\sim0.8$ (the amount of dust already produced in SN 1987A), $0.5$, $0.1$ and $0.5$ 
M$_{\odot}$ for our NGC 604, NGC 595, NGC 588 and NGC 592 models, respectively. For the NGC 588 and NGC 592 models, one can note 
that due to the extreme conditions in the thermalized ISM leading to a rapid sputtering of the grains and a long
interval between SNe, the mass of dust remaining in the cluster as a result of a single injection event becomes negligible 
after a few thousand years. If the dust injection timescale is comparable to the interval 
between supernova explosions, then there is a non-negligible amount of dust present in the cluster during a large fraction of the 
supernova era (which lasts the order of $\sim 4\times 10^7$ years). 

\begin{figure*}[htp]
\plotone{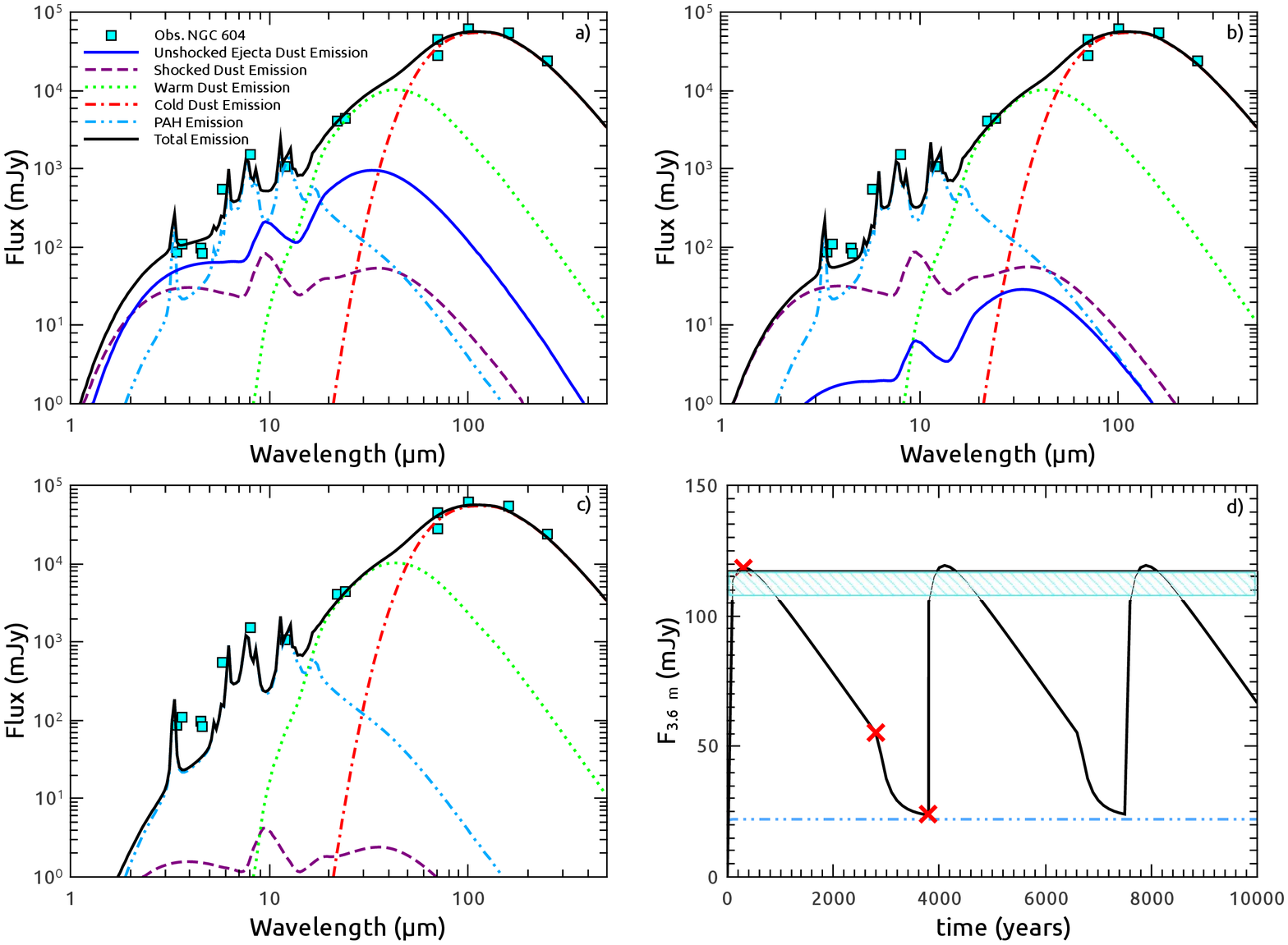}
\caption{The infrared spectral energy distribution for our NGC 604 dust emission model at 300 (panel a), 
2800 (panel b) and 3800 (panel c) years after the first injection event started with an 
interval between events $\sim 3800$ years. The blue-solid and purple-dashed lines depict the emission from dust
residing in the unshocked and shocked ejecta, respectively, for our NGC 604 model. The dash-double dotted blue, dotted-green 
and dash-dotted red lines correspond to the emission of pre-existing PAHs, warm and cold grains in the illuminated H\,{\sc ii} region associated to the 
central cluster, respectively. Total emission is displayed as a black solid line. Observed values (squares 
in cyan) were obtained from the tables provided in \citet{Relanoetal2016}. Panel d (bottom right) presents the evolution
of the flux modeled at 3.6 $\mu$m as compared to the observed value (shaded region in cyan, which includes the associated 
observational error) and the emission from PAHs (dash-dotted light blue line). The red crosses in panel d indicate the times 
displayed in the other panels.}
\label{fig:NGC604}
\end{figure*}

Figure \ref{fig:NGC604} shows the calculated IR SEDs for our NGC 604 model at three different evolutionary times
($300$, $2800$ and $3800$ years) for a total of $0.8$ M$_{\odot}$ of dust condensed in the supernova ejecta. For ease 
of comparison, we show the four separate components required to fit the observations, in a similar fashion as 
\citet{Relanoetal2016}. These evolutionary times were selected because at $\sim 300$ years, the emission at 
$3.6$ $\mu$m from the unshocked and shocked ejecta are approximately equal and thus the total emission is at a local maximum, 
and at $\sim 2800$ years when the incorporation of all the dust into the thermalized medium has been completed. 
Before $\sim 300$ years, the emission arising from dust in the unshocked ejecta is dominant over the emission corresponding to
dust in the thermalized medium. After $\sim 2800$ years, the emission drops fast (as a consequence of rapid grain destruction) 
until $\sim 3800$ years, when the second injection event starts, as shown in panel d in Figure \ref{fig:NGC604}. 

We ought not to omit that the contribution from starlight in the infrared SEDs is orders of magnitude below the observed data 
points and therefore it is not responsible for the infrared excesses we are devoted to explain. At $300$ and $2800$ years after 
the start of the first injection event, the presence of the NIR-MIR excess with respect to the PAHs emission is
evident. Only at $\sim 3800$ years, the excess is marginal. In general the presence of the NIR-MIR excess should be 
detectable for $\sim 85 \%$ of the time spent by the cluster in the supernova era with the assumed supernova rate and heating 
efficiency ($10 \%$). This percentage is calculated from the time during which the dust emission has a clear excess with respect to 
the emission of PAHs, not with the agreement with the observed data of a particular cluster in a particular evolutionary stage. In 
this case, the mass of gas enclosed in the star cluster ($\leq R_{SC}$) is $535$ M$_{\odot}$, $\sim 107$ times more than the typical 
SN ejecta mass we have assumed.

\begin{figure*}[htp]
\plotone{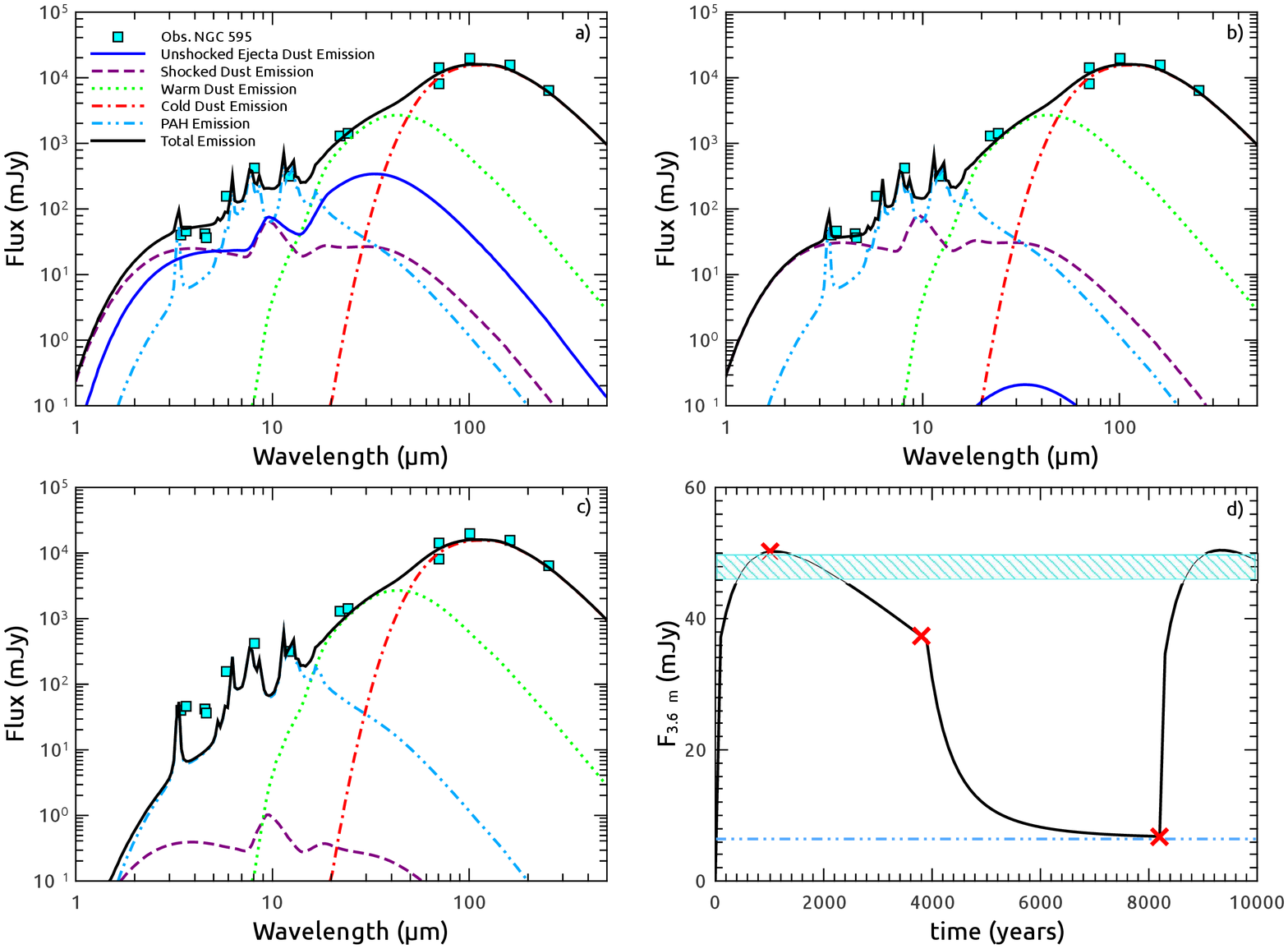}
\caption{Same as Figure \ref{fig:NGC604} but for the model representing NGC 595. In this case, the mass of the
stellar cluster is $2.24\times10^5$ M$_{\odot}$. Relevant evolutionary times in this case are $1000$, $3800$ and
$8200$ years.}
\label{fig:NGC 595}
\end{figure*}

Our calculated IR SEDs for the NGC 595 for the selected evolutionary times are presented in 
Figure \ref{fig:NGC 595}. In this case, a total dust mass of $0.5$ M$_{\odot}$ was used. One can observe that the agreement 
with the observed data points is excellent at $1000$ and $3800$ years, while for $8000$ years the excess with respect to 
the emission of PAHs is absent. However, the NIR-MIR infrared excess associated to these grains would be detectable during 
$\sim 60 \%$ of the evolution of the cluster in this model (see panel d in Figure \ref{fig:NGC 595}). In this model the mass 
of gas in the star cluster is $210$ M$_{\odot}$.

\begin{figure*}[htp]
\plotone{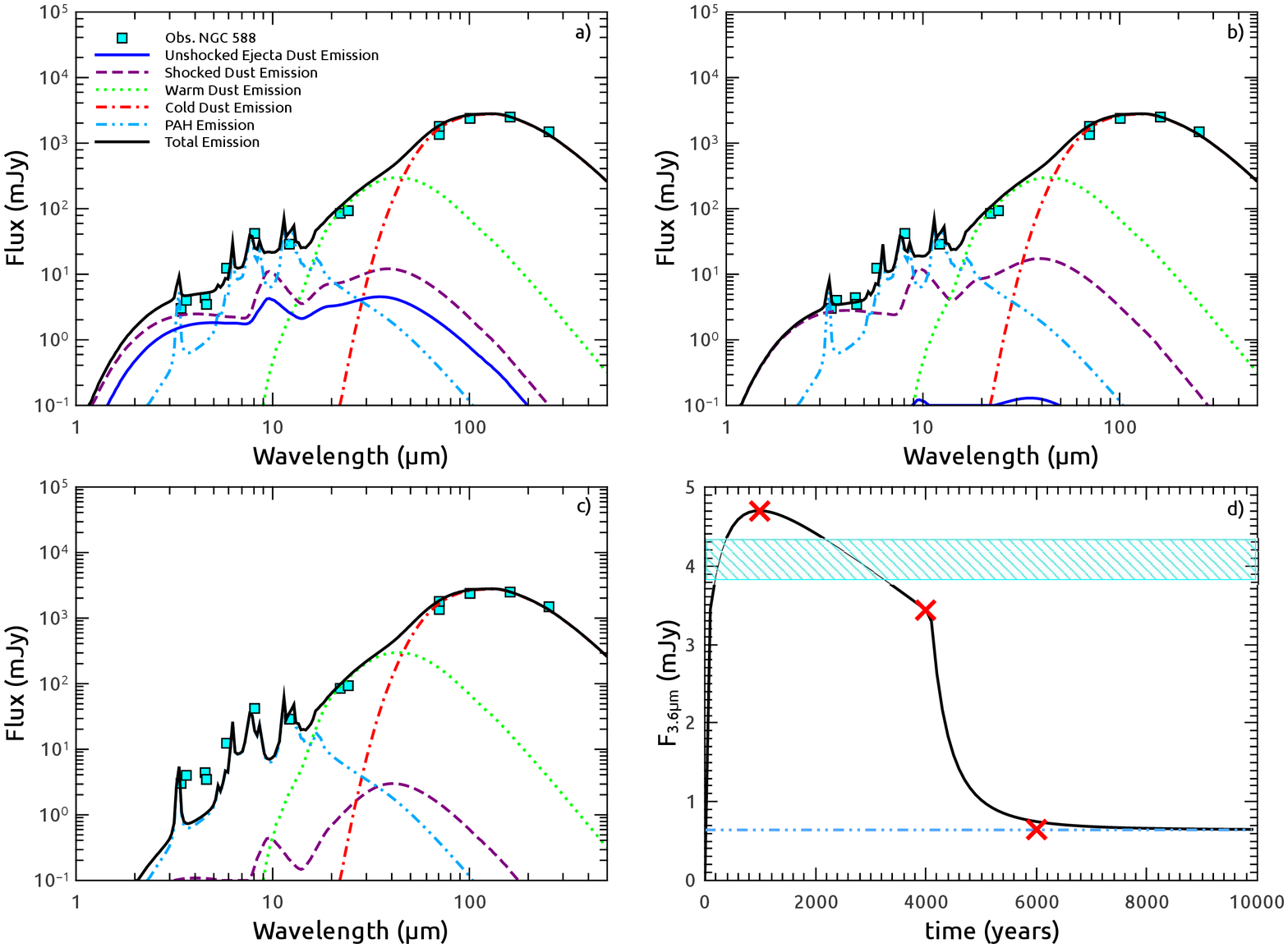}
\caption{Same as Figure \ref{fig:NGC604} but for the model representing NGC 588. In this case, the mass of the
stellar cluster is $6.86\times10^4$ M$_{\odot}$. Panels a, b and c show the calculated fluxes $3.6$ $\mu$m at
$1000$, $4000$ and $10000$ years.}
\label{fig:NGC 588}
\end{figure*}

For our NGC 588 model (evaluated and displayed in Figure \ref{fig:NGC 588}) we used $0.1$ M$_{\odot}$ of injected dust, 
the infrared emission behaves similarly to the previous cases at $1000$ and $4000$ years. At $10000$ years the grain emission 
is overcome by the PAHs emission and the NIR-MIR excess of our interest is absent. Only during $\sim 20 \%$ of the supernova era 
the NIR-MIR excess would be present as shown in the comparison for the modeled and observed flux at $3.6$ $\mu$m
in Figure \ref{fig:NGC 588}. In this case, the enclosed mass of gas in the star cluster is $180$ M$_{\odot}$.

\begin{figure*}[htp]
\plotone{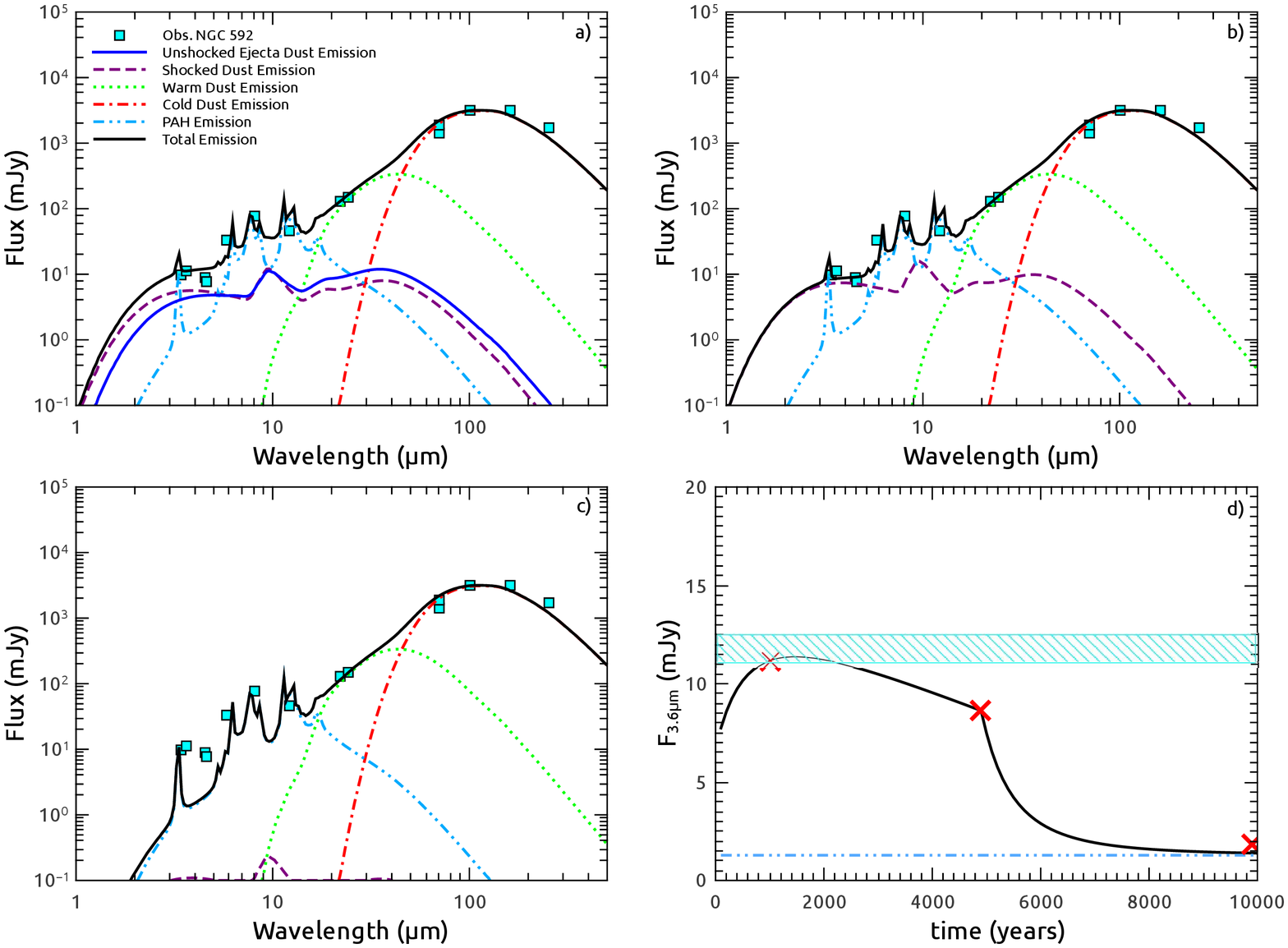}
\caption{Same as Figure \ref{fig:NGC604} but for the model representing NGC 592. In this case, the mass of the
stellar cluster is $3.98\times10^4$ M$_{\odot}$. Panels a, b and c show the calculated flux at
$1000$, $4800$ and $10000$ years.}
\label{fig:NGC 592}
\end{figure*}

In the case of NGC 592 (see Figure \ref{fig:NGC 592}), the results in general are very similar to the case of NGC 588 for 
the three shown evolutionary times even though this cluster is less massive and the assumed amount of dust injected 
within the star cluster is $0.5$ M$_{\odot}$. The stellar mass of NGC 592 is roughly the same as some other prominent clusters 
in M33 like IC 131 and IC 131-West \citep{Relanoetal2010}. For this reason, we will use them as probes for the evolutionary 
trend resultant from the injection and destruction of grains. For our NGC 592, the star cluster contains $106$ M$_{\odot}$ 
of gas.

\begin{figure*}[htp]
\plotone{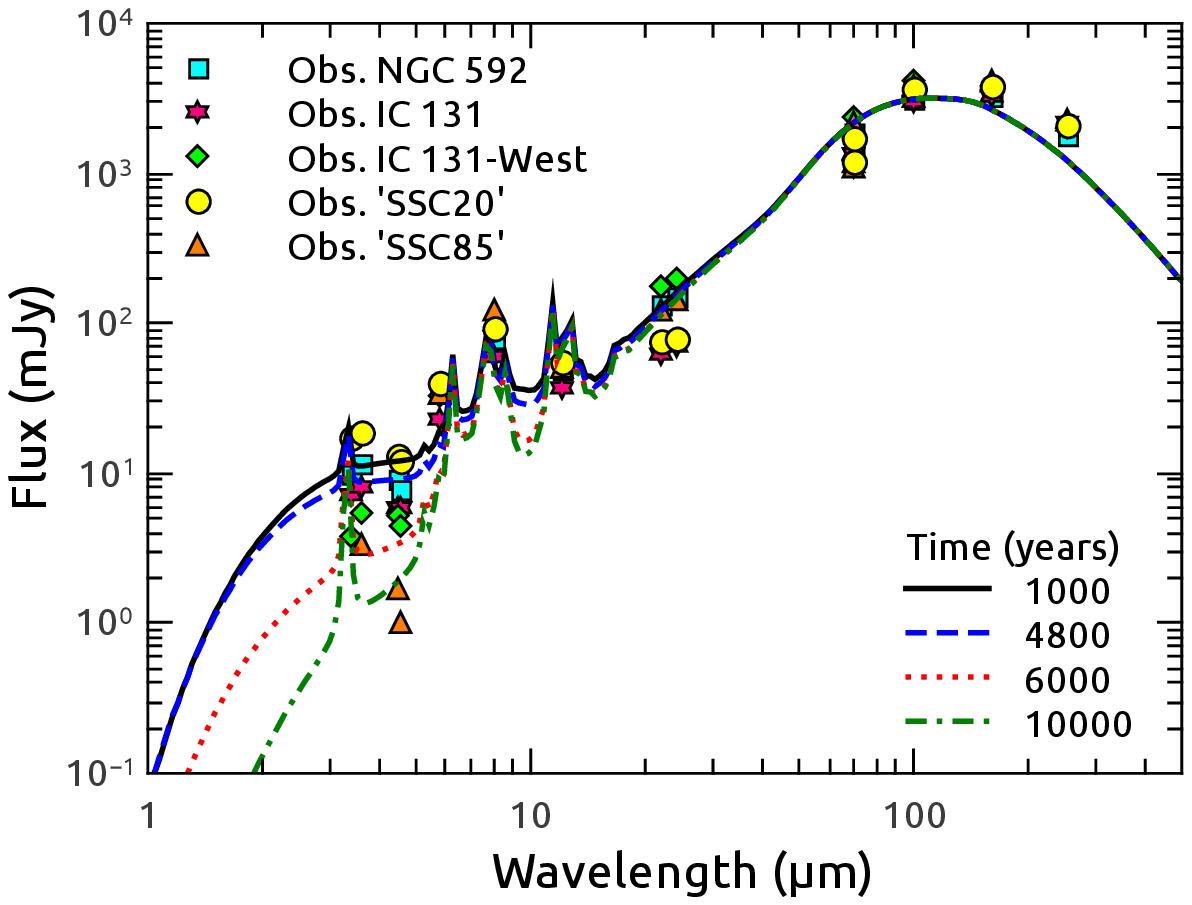}
\caption{Infrared SEDs for regions NGC 592, IC 131, IC 131-West, SSC20 and SSC85. We have overlaid the predictions for our 
NGC 592 model at $1000$ (solid blue line), $4800$ (blue dashed line), $6000$ (red dotted line), and $10000$ 
(green dash-dotted line) years, respectively.}
\label{fig:Clusters}
\end{figure*}

In Figure \ref{fig:Clusters}, the observed IR SEDs for NGC 592 (squares filled with cyan), IC 131 
(pink stars) and ICI 131-West (green diamonds) are shown. Other interesting regions identified by \citet{Relanoetal2013} are 
regions 20 and 85. Region 20 (in the following referred as SSC20) shows strong emission in excess to that of PAHs 
(yellow circles in Figure \ref{fig:Clusters}), whereas the excess is absent for region 85 (referred as SSC85). In the plot, 
we have overlaid our predicted IR SEDs for the NGC 592 model evaluated at $1000$, $4800$, $6000$ and $10000$ years.

In the frame of our models, the differences between these five clusters at short wavelengths ($\leq 10$ $\mu$m) might be 
explained by considering supernova dust injections at different evolutionary stages, off-centered supernovae (evolving in 
the blowout regime as explained by \citet{TenorioTagleetal2015b} and \citet{Silichetal2017}) and/or simply supernovae with different dust yields.

\section{Dust from Other Sources}
\label{sec:sources}

As pointed out earlier in this paper, before the first supernova occurs (at $\sim 3.5$ Myr), 
the intracluster medium is already thermalized by the interaction of the stellar winds of 
individual stars, leading to the launching of a fast outflow, the star cluster wind 
\citep[e.g. ][]{ChevalierClegg1985,Cantoetal2000,Silichetal2004}. Therefore, at the time of the first 
SNe, any residual dust left over from star formation should have been, either destroyed in 
the hot wind or expelled out of the cluster. In the latter case, far from the bulk of the 
stars and in a less hot and less dense medium, the expelled dust must be contributing to 
the emission which is accounted as the warm dust component (green dotted curves in Figures \ref{fig:NGC604} to
\ref{fig:NGC 592}). 

Dust condensed out of the material injected by Wolf-Rayet stars (WR) must also contribute to the 
IR emission in the clusters; however, their contribution to the total dust budget is unclear and most 
likely minor compared to dust injection by SNe \citep[e.g. ][for the case of the Small Magellanic 
Cloud]{Matsuuraetal2013}. Moreover, clusters with a presumably large WR population producing some amount of dust, 
show no NIR-MIR excess (e.g. SSC85), making WR stars unlikely to play a significant role in producing the excess.

\section{Summary}
\label{sec:Summary}

Here we have considered the infrared emission of the dust grains formed in core-collapse SN within young massive
stellar clusters. It is noted that the emission at NIR-MIR wavelengths is more likely to be produced by stochastically-heated 
small grains and that several effects can prevent their fast destruction. As a result, the period during which their emission 
dominates over the foreground PAHs features is enhanced. These effects can be summarized as follows:

\begin{enumerate}

 \item Magnetohydrodynamic turbulence in star clusters \citep{Hirashitaetal2010}, together with the compression of the
 magnetic field lines \citep{Shull1977}, acting in clumpy SN ejecta may enhance the 
 occurrence of grain-grain collisions, producing an excess of small grains \citep{Jonesetal1996}, 
 we thus consider that the grain population is formed only by grains with radii $\lesssim 0.05$ $\mu$m. 

 \item Small grains can actually be traversed by colliding ions if these are energetic enough not to be stopped 
 inside the small dust particles, as shown by \citet{SerraDiazCanoetal2008}. This effect reduces the derived 
 sputtering rates for very small grains thus increasing their lifetimes.

 \item The NIR-MIR infrared excesses are more persistent if the time required to shock-process individual SN ejecta  
 \citep{TangandWang2009} is comparable to the interval between SN explosions.
 
 \item A low efficiency in the thermalization of the kinetic energy of stellar winds and supernova explosions 
 \citep{Silichetal2007,Silichetal2009} would also alleviate the destruction of grains of all sizes.

 \item Star clusters with modest masses ($\sim 10^4$ M$_{\odot}$), if not mass-loaded, may struggle with the capture 
 and interaction of SN ejecta, thus inhibiting the driving of a smooth wind in the sense of the 
 \citet{ChevalierClegg1985}  classical model; hence, the consideration of mass-loaded winds is also crucial in these 
 cases. 
\end{enumerate}

\section{Concluding Remarks}
\label{sec:conclusions}

We have applied our SN dust injection model to the four most prominent star clusters in the M33 galaxy, 
spanning a wide range of masses and dust (from different sources) contents, from the \citet{Relanoetal2013,Relanoetal2016}
sample of H\,{\sc ii} regions and their associated observed infrared SEDs. We found that massive clusters, like NGC 604 and 
NGC 595, should exhibit NIR-MIR excesses during a significant fraction of their evolution, 
especially if the heating efficiency of the thermalized matter is low as suggested by several independent studies 
\citep[e.g.][]{Smithetal2006,Silichetal2009}. With regard to star clusters with masses of a few times $\sim 10^4$ M$_{\odot}$, 
we propose that the evolutionary trends of the NIR-MIR emission obtained from our models is well represented by NGC 592, 
NGC 592, IC 131 and IC 131-West, which have similar masses and their emission at wavelengths $\geq 10$ $\mu$m is almost 
identical. When analyzing other young massive clusters with no available stellar mass estimates, e.g. the star clusters 
with identification numbers 20 and 85 in the \citet{Relanoetal2016} sample, one can observe that while
the emission at wavelengths $\geq 10$ $\mu$m is very similar to that found in NGC 588, NGC 592, IC 131 and IC 131-West,
they differ at short wavelengths by an order of magnitude. We interpret this as evidence that region 20 experienced
grain injection from a supernova in the last few thousand years, whereas region 85 should not have had a recent 
grain injection occurring at its central region.

Our model thus predicts that a NIR-MIR excess might be transiently observed in the spectra of young massive star 
clusters. The characteristic timescale for such an excess to be observed and thus the probability to observe 
it in a sample of the selected clusters depends on the ability to capture the ejecta of individual supernovae inside 
the clusters. In broad terms, the presence of the NIR-MIR excess is an indication of efficient dust 
production by SNe exploding within star clusters. It also suggests the presence of a large population of very small grains 
heated to high temperatures as well as strong grain destruction by ionic collisions.

Further analysis of the implications of the blowout scenario \citep{TenorioTagleetal2015b,Silichetal2017} on 
the survival, manifestation and dispersal of supernova-produced dust grains is left for a future communication.

\acknowledgments

The authors thank G. Tenorio-Tagle, S. Silich and the anonymous Referee for their careful reading and helpful suggestions which greatly 
improved the paper, and M. Rela{\~n}o for answering our questions about her data. Support for this project was provided by the Czech Science 
Foundation grant 209/15/06012S and by project RVO: 6785815. SMG also thanks initial support for the project provided by INAOE and CONACYT, 
M\'exico through research grant 167169. SMG dedicates this work to the memory of his beloved baby And\v el.

\bibliographystyle{apj}
\bibliography{Infrared}

\end{document}